# Heterodyne-detected giant surface optical nonlinearity of transition metal dichalcogenide Dirac semimetal

Syed Mohammed Faizanuddin[1,2,3], Ching-Hang Chien[5], Yao-Jui Chan[1], Chia-Nung Kuo[4], Chin Shuan Lue[4], and Yu-Chieh Wen[1,*]

[1]*Institute of Physics, Academia Sinica, Taipei 11529, Taiwan*

[2]*Nano Science and Technology Program, Taiwan International Graduate Program, Academia Sinica, Taipei 11529, Taiwan*

[3]*Department of Engineering and System Science, National Tsing Hua University, Hsinchu 300044, Taiwan*

[4]*Department of Physics, National Cheng Kung University, Tainan 701401, Taiwan*

[5]*Research Center for Applied Sciences, Academia Sinica, Taipei 11529, Taiwan*

*Correspondence to: ycwen@phys.sinica.edu.tw (Y.C.W.)*

**Abstract:**

Three-dimensional topological semimetals were shown to host anomalously strong optical nonlinearity, but yet whether such enhancement can occur at their surfaces via topological boundary states remains an open question. Here we tackle this question by examining polarization-dependent rotational-anisotropy optical second harmonic generation (SHG) from transition metal dichalcogenide Dirac semimetal PdTe$_2$. We find its reflected SHG to follow $C_{3v}$ surface symmetry with a time-varying intensity dictated by oxidation kinetics of the material after its surface cleavage, confirming the surface-derived origin of SHG. The complex surface nonlinear optical susceptibility tensor is fully characterized by heterodyne interferometry, unveiling a hidden giant out-of-plane response $\chi^{(2)}_{zzz}$ (> 10.6 nm$^2$/V or, equivalently, an effective bulk nonlinear susceptibility > 10.3 nm/V). Our results highlight the topological surfaces/interfaces as a new route toward nonlinear optical applications with released symmetry constraints, and demonstrate SHG as a viable means for *in situ* probing of surface chemistry of topological systems.

**Keywords**

Transition metal dichalcogenide, Dirac semimetal, phase–resolved second harmonic generation, surface nonlinear optics, topological surface states.

Realizing various topological phases in three-dimensional condensed matters has offered unique opportunities for exploring exotic properties of Dirac and Weyl fermions [1, 2]. Chiral anomaly [1, 2], axion electrodynamics [3], and surface Fermi arcs [4] are just a few recently known examples. Nonlinear optical responses garnered particular attentions for their close correlation to band topology and quantum geometric properties of electron Bloch wavefunctions [5-9]. Rich unusual phenomena were proposed and/or observed, ranging from quantized circular photogalvanic effect [5, 6], enhanced shift/injection currents [6-8], record-high optical second harmonic generation (SHG) [10], to nonlinear Hall effect in the microwave regime [9]. Beyond the bulk, peculiar topological surface states potentially host unconventional nonlinear optical signatures that would enable new applications with released symmetry constraints [11, 12]. However, surface nonlinear optics of topological semimetals is less explored due to experimental difficulties in separating the surface from bulk responses [12-15].

Dirac semimetals with preserved inversion symmetry offer a platform for investigating the topological surface nonlinear optics due to the absence of the bulk electric-dipole contribution to the even-order nonlinearity. Recent studies on anisotropic shift currents in transition-metal dichalcogenide (Pd, Ni)Te$_2$ [16, 17] and SHG from nodal-line semimetals ZrSiS and HfGeTe [14, 15] have led to remarkable development of high-sensitivity broadband photodetection and efficient frequency conversion, respectively. Surface specificity of these effects is, however, not necessarily true, even with bulk inversion symmetry, because electric-quadrupole and magnetic-dipole contributions from the rapid field variation over the bulk inevitably come to play [13, 18, 19]. A unique fingerprint of the surface second-order nonlinearity has yet to be found experimentally. Whether it can be enhanced in topological systems remains an open question. Here we address this issue by examining the correlation between optical SHG and surface reactions.

Another ubiquitous difficulty comes from the technical paucity in characterizing the anisotropic complex-valued surface nonlinear optical susceptibility tensor, $\overleftrightarrow{\chi}^{(2)}$. In Dirac semimetals with rotational invariance, protected Dirac cones are usually formed through a band inversion on the rotation axis (labelled along $\hat{z}$). This axis is often an easy direction one creates a stable atomically-flat surface for many materials [14-17, 20], suggesting that for these surfaces, intriguing topological signatures would be encoded in the out-of-plane anisotropy of $\overleftrightarrow{\chi}^{(2)}$. However, determination of its purely out-of-plane $zzz$ element is difficult because of its inherent interference with other elements with identical in-plane rotational isotropy [20, 21]. In the previous (homodyne-detected) SHG studies on Dirac semimetals [14, 15], the phase differences between the tensor elements were simply ignored in approximating $\left|\chi_{zzz}^{(2)}\right|$. None of $\chi_{zzz}^{(2)}$ can be retrieved with nontrivial quantitative certainty.

In this Letter, we present a heterodyne-detected (HD-) SHG spectroscopic study on transition-metal dichalcogenide 1T-PdTe$_2$. It serves as a prototype Dirac semimetal with its topological characters naturally formed in a single Te $p$-orbital manifold under a trigonal crystal field with large inter-layer hopping [22, 23]. We confirm the surface origin of its SHG by realizing that the SHG follows $C_{3v}$ surface symmetry of the crystal with a time-varying intensity dictated by the surface oxidation kinetics. Moreover, the complex $\overleftrightarrow{\chi}^{(2)}$ is fully characterized via the heterodyne interferometric scheme, through which we evidence a hidden giant out-of-plane $\chi_{zzz}^{(2)}$ of the surface. Our study highlights the potential of topological surfaces/interfaces for nonlinear optical applications with enhanced responses and released symmetry constraints.

The sample studied is a (001) 1T-PdTe$_2$ single crystal grown by the self-flux method. Detailed growth conditions and structural and electronic characterizations have been reported elsewhere [16, 24]. Our HD-SHG experiment was conducted with a fundamental field $E(\omega)$ at 1.55 eV in the ambient condition. As sketched in Figure 1a, the input field, after passing through a *y*-cut quartz acting as a local oscillator (LO) and a 2mm-thick SF11 glass being a phase modulator (PM), was focused onto the sample surface with the fluence of 2 mJ/cm$^2$ and an incident angle of 45°. The SH field radiated from the sample in reflection, $E_s(2\omega)$, was proportional to the effective surface nonlinear polarization, as given by $E_s(2\omega) \propto \chi_{eff} E(\omega) E(\omega)$ with $\chi_{eff}$ the effective surface nonlinear optical susceptibility. $E_s(2\omega)$ interfered with the LO field reflected from the sample. The resultant SHG spectral interferogram was measured by a polychromator (Figure 1b) and Fourier-analyzed with normalization against a *z*-cut quartz reference to deduce complex $\chi_{eff}$ of the sample in MKS units [25]. (See the Supplementary Information (SI) section S1 and Ref. [25] for details and accuracy validation of the heterodyne analysis, respectively, and also see SI section S4 for validity of our PdTe$_2$ measurement justified through homogeneity and optical damage tests and the quadratic power dependence of the SHG intensity.)

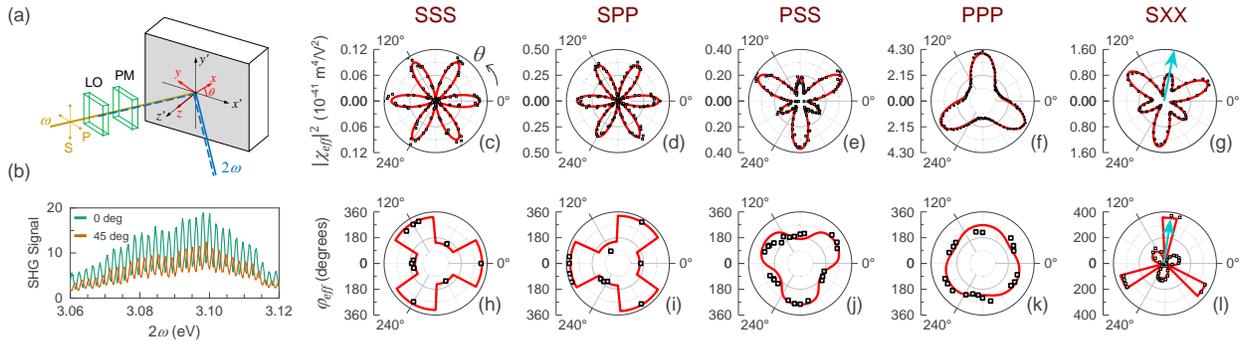

**Figure 1.** Heterodyne-detected (HD-) SHG from PdTe$_2$ single crystal at the steady state. (a) Schematics of the HD-SHG setup. LO and PM denote the local oscillator and the phase modulator, respectively. The $(x', y', z')$ lab coordinates set with $\hat{z}'$ along the surface normal and $\hat{x}' - \hat{z}'$ being the scattering plane relate to the $(x, y, z)$ crystal coordinates via $\hat{z}' \parallel \hat{z}$ and the azimuthal angle $\theta$ between $\hat{x}'$ and $\hat{x}$. (b) Representative SHG spectral interferograms measured by SPP-polarized

HD-SHG, where a clear ~$\pi$ phase shift is seen at two selected $\theta$. (c)-(g) $|\chi_{eff}(\theta)|^2$ and (h)-(l) $\varphi_{eff}(\theta)$ measured by HD-SHG for various polarization combinations (dots), in comparison with theoretical fits (lines). An arrow in (g) and (l) marks a high-symmetry axis of the SXX pattern, which rotates away from the mirror plane of the $D_{3d}$ crystal.

Figure 1c-1l shows the measured complex $\chi_{eff}$ of PdTe$_2$ for various polarizations as a function of the azimuthal angle $\theta$ between the crystal axis $\hat{x}$ ($\parallel [11\bar{2}0]$) and the optical scattering plane. We label the polarizations of SHG and two fundamental photons in order as S-, P-polarization, or "X" for linear polarization at 45º away from the scattering plane. (Later, a modified homodyne setup with two cross-polarized input beams was employed to access $|\chi_{eff}|$ with SSP polarization.) Anisotropic material responses are readily seen through the SHG intensity, as revealed by $|\chi_{eff}(\theta)|^2$ in Figure 1c-1g, as well as the phase of $\chi_{eff}(\theta)$ (denoted as $\varphi_{eff}$) in Figure 1h-1l. The $|\chi_{eff}(\theta)|^2$ patterns exhibit pronounced six-fold and three-fold symmetry for SSS/SPP and PSS/PPP/SXX polarizations, respectively. But, more clearly, inclusion of the measured $\varphi_{eff}(\theta)$ clarifies the three-fold nature of $\chi_{eff}(\theta)$ for all the polarization combinations.

Analyses of the deduced $\chi_{eff}(\theta)$ must rely on identification of its surface- versus bulk-derived origin. It is realized here by inspecting the relation of the SHG to surface oxidation. It was known from X-ray photoemission studies that surface oxidation of pristine PdTe$_2$ forms a robust sub-nanometric TeO$_2$ skin layer within 30 minutes of its exposure to air [26, 27]. Surface SHG is thus expected to change accordingly, but the bulk SHG via electric quadrupoles and magnetic dipoles should remain the same. Figure 2 shows the rotational-anisotropy SHG intensities, taken ~6 min, 3 hours, and 5 days after mechanical cleavage of the crystal along the (001) plane. (The 5-day data taken from Figure 1 represent the steady-state results.) Indeed, the magnitudes of those lobes are found to vary with time after the cleavage for all the polarizations, with the changes up to >300 % (e.g., PSS at $\theta = 30°$). More remarkably, the PPP, PSS, and SSP patterns display geometric changes. These observations confirm that the SHG from centrosymmetric PdTe$_2$ is surface sensitive and, therefore, the responsible contribution from the surface electric dipoles. Interestingly, although the SHG modification occurs mainly within the first 3 hours, the PPP pattern keeps evolving in shape on a longer timescale (Figure 2j). It reveals that a slow surface process takes place in addition to the known oxidation kinetics, as discussed later.

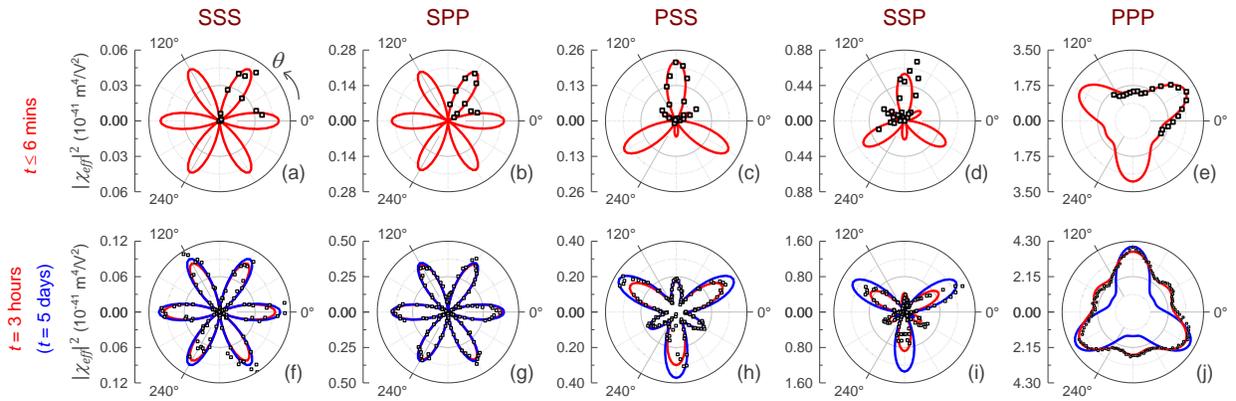

**Figure 2.** Rotational-anisotropy SHG from PdTe$_2$ surface for different polarizations at various time delays $t$ since its surface cleavage in ambient air. The measured $\left|\chi_{eff}(\theta)\right|^2$ is given for (a)-(e) $t \leq 6$ mins and (f)-(j) $t \approx 3$ hours (dots), in comparison with theoretical fits (red lines). Blue curves in (f)-(j) are fits to the measurement results at the steady state ($t \approx 5$ days). In (a)-(e), the fewer data within a limited angular range that covers the extrema of $C_{3v}$–dictated SHG lobes were acquired for improving the time resolution.

With the surface origin identified, we can then relate $\chi_{eff}$ to $\overleftrightarrow{\chi}^{(2)}$ of the surface under the electric-dipole approximation. For PdTe$_2$ having a layered CdI$_2$-type structure of the trigonal space group $P\bar{3}m1$ [26], the point group symmetry reduces from $D_{3d}$ in the bulk to $C_{3v}$ at the truncated (001) surface. This activates the surface SHG via four independent non-zero $\overleftrightarrow{\chi}^{(2)}$ elements, as given by: $\chi^{(2)}_{yyy} = -\chi^{(2)}_{yxx} = -\chi^{(2)}_{xxy} = -\chi^{(2)}_{xyx}$, $\chi^{(2)}_{xzx} = \chi^{(2)}_{yzy} = \chi^{(2)}_{xxz} = \chi^{(2)}_{yyz}$, $\chi^{(2)}_{zxx} = \chi^{(2)}_{zyy}$, and $\chi^{(2)}_{zzz}$. They constitute $\chi_{eff}$ for a given optical configuration via the relation [13, 28]: $\chi_{eff} = [\hat{e}_{SH} \cdot \overleftrightarrow{L}_{SH}] \cdot \overleftrightarrow{\chi}^{(2)} : [\hat{e}_1 \cdot \overleftrightarrow{L}_1][\hat{e}_2 \cdot \overleftrightarrow{L}_2]$, where $\overleftrightarrow{L}_n$ and $\hat{e}_n$ are the tensorial Fresnel transmission coefficient and the polarization unit vector, respectively, for SH and two involved fundamental fields (denoted by $n = SH$, 1, 2, respectively). $\chi_{eff}$ for the adopted polarizations is expressed as

$$\chi_{eff}^{SSS} = \cos(3\theta) \cdot c_1 \chi^{(2)}_{yyy}, \tag{1a}$$

$$\chi_{eff}^{SPP} = \cos(3\theta) \cdot c_2 \chi^{(2)}_{yyy}, \tag{1b}$$

$$\chi_{eff}^{PSS} = c_3 \chi^{(2)}_{zxx} + \sin(3\theta) \cdot c_4 \chi^{(2)}_{yyy}, \tag{1c}$$

$$\chi_{eff}^{SSP} = c_5 \chi^{(2)}_{xxz} + \sin(3\theta) \cdot c_6 \chi^{(2)}_{yyy}, \tag{1d}$$

$$\chi_{eff}^{SXX} = c_7 \chi^{(2)}_{xxz} + [\sin(3\theta) \cdot c_8 + \cos(3\theta) \cdot c_9] \chi^{(2)}_{yyy}, \tag{1e}$$

$$\chi_{eff}^{PPP} = \chi^{(2)}_{iso} + \sin(3\theta) \cdot c_{10} \chi^{(2)}_{yyy}, \tag{1f}$$

with $\chi^{(2)}_{iso} \equiv c_{11} \cdot \chi^{(2)}_{zxx} + c_{12} \cdot \chi^{(2)}_{xxz} + c_{13} \cdot \chi^{(2)}_{zzz}$. (1g)

We have lumped $\overleftrightarrow{L}$ and polarization projection factors into constants $c_i$ ($i = 1\sim 13$). (See SI section S2 for their expressions.) Clearly, $\chi_{eff}$ is composed of an anisotropic in-plane *yyy* term plus isotropic contributions from the out-of-plane tensor elements. Characteristics of $\chi_{eff}$ for SSS versus SPP, and SSP versus SXX, are effectively the same.

For experimental analyses, we calculate $\overleftrightarrow{L}$ and $c_i$ from the linear dielectric constant of PdTe$_2$ measured by ellipsometry, and then use Eq. (1) to fit the measured $\chi_{eff}(\theta)$ for the different polarizations simultaneously with the estimated $c_i$. (See SI section S2 for the ellipsometric results and calculations of $\overleftrightarrow{L}$ and $c_i$.) As shown in Figure 1c-1l, the complex $\chi_{eff}(\theta)$ of the oxidized surface at the steady state, including $\chi_{eff}^{SXX}(\theta)$ with rotated high-symmetry axes (as marked by arrows in Figure 1g, 1l), can be fitted satisfactorily well using the $C_{3v}$–dictated $\overleftrightarrow{\chi}^{(2)}$. This yields $\chi_{ijk}^{(2)} = -1.3 + 3.5i, 6.0 - 6.6i, -3.5 + 6.9i$, and $-94 - 422i$ (in units of $10^{-20}$ m$^2$/V) for $ijk =$ *yyy*, *xxz*, *zxx*, and *zzz*, respectively. Now, with the oxidized surface known to have $C_{3v}$ symmetry as the pristine case, one may expect that the surface symmetry is preserved throughout the surface processes after its cleavage. Indeed, it is supported by an extension of our valid analysis to the SHG patterns measured at short time delays *t* through $C_{3v}$-derived Eq. (1), as revealed in Figure 2, although fewer data were acquired within $t < 6$ mins.

More quantitatively, we characterize the time evolution of $\overleftrightarrow{\chi}^{(2)}$ quantities through, without loss of generality, the SPP-polarized HD-SHG for $\chi_{yyy}^{(2)}(t)$ and the PSS-, SSP-, and PPP-polarized rotational-anisotropy SHG intensities. For the latter, fitting the $|\chi_{eff}(\theta)|^2$ data alone determines the magnitude of each term in Eq. (1c) - (1f) and their absolute phase difference for a given *t*. The sign uncertainty of the phase relation can be resolved with help of the measured steady-state complex $\overleftrightarrow{\chi}^{(2)}$ by assuming a continuous phase evolution. Together with the phase of $\chi_{yyy}^{(2)}(t)$ measured, one can readily retrieve the time-dependent complex $\chi_{zxx}^{(2)}, \chi_{xxz}^{(2)}$, and $\chi_{iso}^{(2)}$ from the rotational-anisotropy SHG intensities.

Figure 3 shows representative transient SHG data acquired by a quick sampling of the extrema of the $C_{3v}$–dictated SHG lobes at $\theta = m\pi/6$ with $m$ an integer. These raw data reveal rich temporal characteristics, including profound non-monotonic changes of $|\chi_{eff}^{PPP}|^2$ on multiple timescales and a monotonic decay of the phase of $\chi_{eff}^{SPP}$ for $t < 2$ hours. (See Fig. S4 in SI for $|\chi_{eff}(t)|^2$ with other polarizations.) With the above, the full complex $\overleftrightarrow{\chi}^{(2)}$ tensor is deduced as a function of time delays after cleavage, as shown in Figure 4. Each of the deduced elements displays a simple increase/decrease with time for $t < 2$ hours, in contrast to the non-monotonic $|\chi_{eff}(t)|^2$ owing to mixing of the real and imaginary parts (e.g., Figure 3a). $\chi_{yyy}^{(2)}$, $\chi_{zxx}^{(2)}$, and $\chi_{xxz}^{(2)}$ become stable afterward. But, $\chi_{iso}^{(2)}$ exhibits an additional slow kinetics through $\chi_{zzz}^{(2)}$ with a timescale of tens of hours, as the cause of the slowly evolving PPP pattern observed in Figure 2j and 3a.

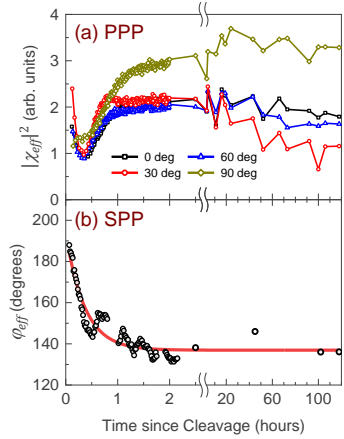

**Figure 3.** Time evolution of $|\chi_{eff}|^2$ and $\varphi_{eff}$ of PdTe$_2$ crystal after its surface cleavage. (a) Measured PPP-polarized $|\chi_{eff}|^2$ at selected $\theta$ and (b) phase of SPP-polarized $\chi_{eff}$ at $\theta = 0°$ (dots). In (b), an oscillatory artifact with a period of 15 min was caused by instability of our environmental temperature. It is removed later by post-data processing, as shown by a line in (b).

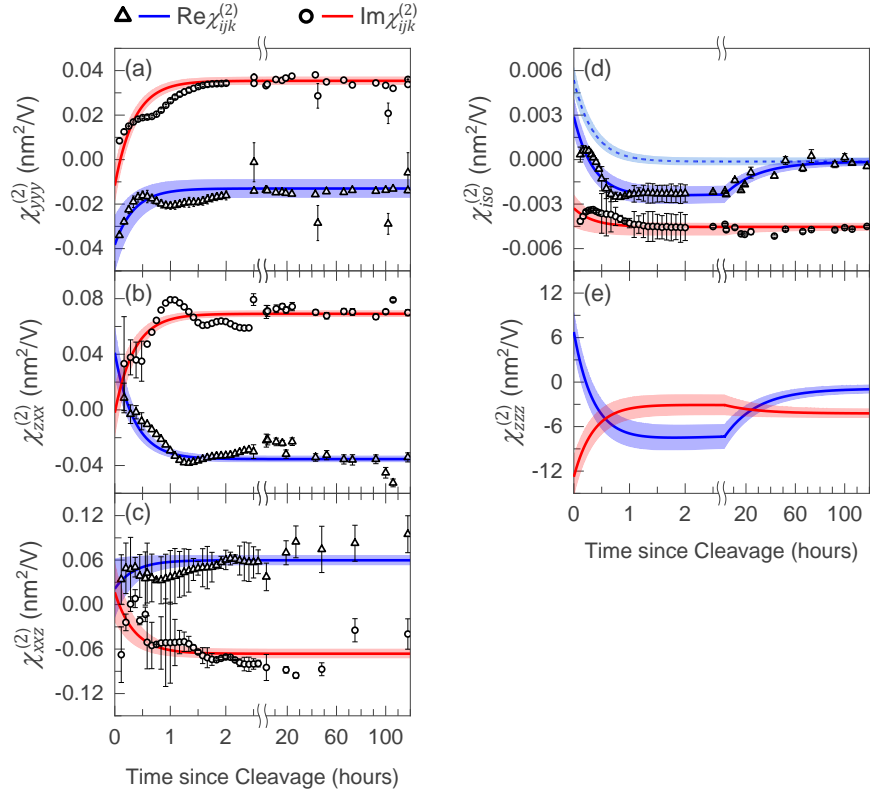

**Figure 4.** Complex surface nonlinear optical susceptibility $\overleftrightarrow{\chi}^{(2)}$ of (001) PdTe$_2$ as a function of time delays after the surface cleavage. Dots are the measured data for (a) *yyy*, (b) *zxx*, (c) *xxz* elements, and (d) $\chi^{(2)}_{iso}$, and solid lines are the corresponding exponential fits using (a)-(c) a single time constant (0.35 hour) or (d) two time constants (0.35 and 26 hours). In (d), a blue dashed line is the fitting curve for $\mathrm{Re}\chi^{(2)}_{iso}(t)$ with the slow exponential term excluded, revealing the necessity of this term. (e) $\chi^{(2)}_{zzz}(t)$ deduced from the fits in (b)-(d) via Eq. (1g). Shadow regions in (a) – (e) represent the fitting uncertainty.

Our observations can be described by a phenomenological model based on exponential decays: $\chi_{ijk}^{(2)}(t) = \chi_{ijk}^{SS} + \sum_n \Delta\chi_{ijk}^n \exp(-t/\tau_n)$, where $\chi_{ijk}^{SS}$ and $\Delta\chi_{ijk}^n$ are complex amplitudes, $\tau_n$ is a time constant, and $n = 1, 2$ (1) for *iso* (*yyy*, *zxx*, and *xxz*) terms, as shown by the fits in Figure 4. We find $\tau_1 = 0.35 \pm 0.1$ hour and $\tau_2 = 26.3 \pm 6.8$ hours. As expected, the deduced $\tau_1$ agrees well with the timescale of the surface oxidation kinetics observed in the photoemission studies [26, 27]. However, the origin of the slow $\tau_2$ kinetics remains elusive. It cannot be explained by atomic reconstructions ($t \ll \tau_1$) or a secondary surface reaction (precluded by the photoemission spectra [26, 27]). We argue that this slow process is probably driven by surface-to-bulk diffusion of defects, which modifies $\chi_{zzz}^{(2)}$ through varying surface carrier concentrations at the diffusion rate [16, 20]. For substantiation, one may use the Arrhenius equation and the observed $\tau_2^{-1}$ (approximated as the jump frequency of diffusion) to estimate the activation energy $E_A$ of defect diffusion [29], and find $E_A$ ~1.02 eV in agreement with the density functional theory (DFT)-derived $E_A$ (~1.19 eV) for Te vacancies diffusing from the surface to the sub-surface Te atomic layer in the outmost unit cell [30].

The purely out-of-plane $\chi_{zzz}^{(2)}$ resolved from $\chi_{iso}^{(2)}$ emerges to be the leading element among $\overleftrightarrow{\chi}^{(2)}$. One can get an intuition about its dominant role by noticing a much isotropic and stronger PPP rotational-anisotropy SHG patterns compared to the other polarization combinations (Figure 2). It is so even if the contribution of $\chi_{zzz}^{(2)}$ has been suppressed by a small $c_{13}$ ($< c_{1\sim12}$) due to strong screening of the out-of-plane electric fields in metallic $PdTe_2$. Note that $\left|\chi_{zzz}^{(2)}\right| = 4.32 \pm 0.46$ nm$^2$/V (for the oxidized surface) is comparable to those of the existing record-high materials [10, 14, 15, 31-33], such as two-dimensional material $MoS_2$, $WS_2$, and $Ge_2Se_2$ monolayers with $\chi_{ijk}^{(2)}$ of 0.8, 5.8, and 6 nm$^2$/V, respectively [31]. To compare $PdTe_2$ surface with three-dimensional materials, we convert $\overleftrightarrow{\chi}^{(2)}$ to an effective bulk nonlinear optical susceptibility $\overleftrightarrow{\chi}_B^{(2)}$ via $\overleftrightarrow{\chi}_B^{(2)} \approx \overleftrightarrow{\chi}^{(2)}/d$ with $d$ the thickness of the effective surface layer with broken inversion symmetry. One would use DFT calculations to find $d \approx 10.3$ Å [23]. Alternatively, knowing that the surface atomic layers which can oxidize affect the SHG sensitively, one may approximate $d$ for the pristine surface as the thickness of TeO$_2$ skin layer on an oxidized PdTe$_2$ ($\approx 8.8$ Å [26]). These estimates of $d$ yield $\left|\chi_{B,zzz}^{(2)}\right| = 3.75 \sim 5.43$ nm/V at the steady state. This value is much higher than zinc-blende semiconductors and ferroelectric oxides [32, 33] and close to the record-high Weyl semimetal TaAs (7.2 nm/V [10]). Finally, we point out that the giant $\chi_{zzz}^{(2)}$ of the PdTe$_2$ surface is qualitatively robust. It does not originate from the TeO$_2$ skin layer because its formation weakens $\chi_{zzz}^{(2)}$ (Figure 4), suggesting an even higher quantity before surface oxidation ($\left|\chi_{zzz}^{(2)}\right| = 14.5 \pm 3.9$ nm$^2$/V at $t = 0$ min estimated by extrapolation).

In summary, we have investigated second harmonic generation of type-II Dirac semimetal PdTe$_2$ and confirmed its surface-derived origin by inspecting that the SHG followed the surface $C_{3v}$ symmetry and oxidation kinetics to vary profoundly after the surface cleavage. The full complex $\overleftrightarrow{\chi}^{(2)}$ tensor was resolved by the heterodyne-detected scheme, through which we evidenced a hidden giant out-of-plane $\chi_{zzz}^{(2)}$. Our study therefore supports the topological interfaces as a new route toward nonlinear optical applications with large responses and released symmetry constraints. Moreover, this work highlights the heterodyne scheme for resolving $\overleftrightarrow{\chi}^{(2)}$ elements with identical rotational anisotropy, which generally includes surface/bulk electric-dipole, electric quadrupole, and magnetic dipole responses. What we develop here therefore enhances the spectroscopic perception to symmetries of different degrees of freedom in condensed matters.

- **ASSOCIATED CONTENT**

**Supporting Information**

Analysis of HD-SHG spectral interferogram; ellipsometric measurement and Fresnel coefficients; effective surface nonlinear optical susceptibility of quartz reference; supplementary figures.

- **AUTHOR INFORMATION**

**Corresponding Author**

**Yu-Chieh Wen** – *Institute of Physics, Academia Sinica, Taipei 11529, Taiwan*; orcid.org/0000-0001-7124-2355; Email: ycwen@phys.sinica.edu.tw

**Authors**

**Syed Mohammed Faizanuddin** – *Institute of Physics, Academia Sinica, Taipei 11529, Taiwan; Nano Science and Technology Program, Taiwan International Graduate Program, Academia Sinica, Taipei 11529, Taiwan; Department of Engineering and System Science, National Tsing Hua University, Hsinchu 300044, Taiwan*

**Ching-Hang Chien** – *Research Center for Applied Sciences, Academia Sinica, Taipei 11529, Taiwan*

**Yao-Jui Chan** – *Institute of Physics, Academia Sinica, Taipei 11529, Taiwan*

**Chia-Nung Kuo** – *Department of Physics, National Cheng Kung University, Tainan 701401, Taiwan*

**Chin Shuan Lue** – *Department of Physics, National Cheng Kung University, Tainan 701401, Taiwan*

**Notes**

The authors declare that they have no competing interest.

**Acknowledgments**

The authors thank Si-Tong Liu and Tien-Dat Tran for experimental supports and David Hsieh, Yia-Chung Chang, and Long Ju for helpful discussions. This work was funded by Academia Sinica (grant number: AS-iMATE-113-12, AS-iMATE-113-14, AS-GCS-111-M01, and AS-iMATE-109-14), and National Science and Technology Council, Taiwan (grant number: NSTC 113-2112-M-001-055-).

Supplementary Information for

# Heterodyne-detected giant surface optical nonlinearity of transition metal dichalcogenide Dirac semimetal

Syed Mohammed Faizanuddin[1,2,3], Ching-Hang Chien[5], Yao-Jui Chan[1], Chia-Nung Kuo[4], Chin Shuan Lue[4], and Yu-Chieh Wen[1,*]

[1]Institute of Physics, Academia Sinica, Taipei 11529, Taiwan

[2]Nano Science and Technology Program, Taiwan International Graduate Program, Academia Sinica, Taipei 11529, Taiwan

[3]Department of Engineering and System Science, National Tsing Hua University, Hsinchu 300044, Taiwan

[4]Department of Physics, National Cheng Kung University, Tainan 701401, Taiwan

[5]Research Center for Applied Sciences, Academia Sinica, Taipei 11529, Taiwan

*Correspondence to: ycwen@phys.sinica.edu.tw (Y.C.W.)

**S1. Analysis of HD-SHG spectral interferogram**

**S2. Ellipsometric measurement and Fresnel coefficients**

**S3. Effective surface nonlinear optical susceptibility of quartz reference**

**S4. Supplementary figures**

## S1. Analysis of HD-SHG spectral interferogram

We followed Ref [1] to analyze the SHG interferogram for deduce $\chi_{eff}$ of the PdTe$_2$ crystal, as described in details below.

In the HD-SHG experiments, we measured the SHG spectral interferogram, $S_q(2\omega)$, created by SH field radiated from the sample in reflection $E_q(2\omega)$ and the LO field $E_{LO}(2\omega)$ reflected from the sample. [Subscript $q$ (= $s$ or $ref$) used here and other symbols below refers to the PdTe$_2$ sample and reference $z$-cut quartz, respectively.] $S_q(2\omega)$ is affected by a relative time delay, $\Delta T$, between the fundamental and LO pulses created by the phase modulator, as expressed by

$$S_q(2\omega) \propto \left|E_q(2\omega) + r_q(2\omega) \cdot E_{LO}(2\omega) \cdot e^{i2\omega\Delta T}\right|^2 \quad (S1)$$

$$= \left|E_q(2\omega)\right|^2 + \left|r_q(2\omega) \cdot E_{LO}(2\omega)\right|^2 + E_q(2\omega) \cdot r_q^*(2\omega) \cdot E_{LO}^*(2\omega) \cdot e^{-i2\omega\Delta T} + c.c.,$$

where $r_q(2\omega)$ is the reflection coefficient, and * denotes complex conjugates. We conduct Fourier transform of $S_q(2\omega)$ to yield $\tilde{S}_q(\tau)$, such that the two dc terms and the two cross terms in Eq. (S1) are separated at $\tau = 0$ and $\pm\Delta T$. We extract one cross term, $E_q(2\omega) \cdot r_q^*(2\omega) \cdot E_{LO}^*(2\omega) \cdot e^{-i2\omega\Delta T}$, and deduce its ratio $R(2\omega)$ for PdTe$_2$ with respect to the quartz. With $E_q(2\omega) \propto [\chi_{eff}]_q E(\omega)E(\omega)$, we can use the measured $R(2\omega)$ to determine $[\chi_{eff}]_s$ with $r_q(2\omega)$ and $[\chi_{eff}]_{ref}$ estimated from the known linear dielectric constants and nonlinear susceptibility of quartz. (See SI section S2 and S3 for details of the estimations, respectively.)

In addition to the HD-SHG analysis, we utilized the same optical setup, but with the LO crystal removed, to measure $|\chi_{eff}(\theta)|^2$ through rotational-anisotropy SHG intensity with SSS, SPP, PSS, PPP, and SXX polarization combinations. To access $|\chi_{eff}|^2$ with the SSP polarization,

we modify the setup to have two cross-polarized input beams with their incident angles of $\beta_{F1} = 42.5°$ and $\beta_{F2} = 47.5°$. All the measurements were based on a 1-kHz Ti:sapphire amplifier with a pulse duration of ~60 fs at 1.55 eV.

## S2. Ellipsometric measurement and Fresnel coefficients

We used an ellipsometer (VASE® by J.A. Wollam) to measure the linear dielectric function, $\varepsilon(E)$, of the PdTe$_2$ crystal at photon energy $E$ between 0.62 ~ 4.12 eV. The setup configuration can be found elsewhere [2]. Fig. S1(a) shows the ellipsometry parameters, Ψ and Δ, measured at different angles of incidence. To deduce the complex $\varepsilon(E)$ from them with the Kramers-Kronig relation obeyed, we fit the measured $\Psi(E)$ and $\Delta(E)$ with $\varepsilon(E)$ described by a Drude–Lorentz model [2, 3]:

$$\varepsilon(E) = \varepsilon_\infty - \frac{A_D B_D}{E^2 + iB_D E} + \frac{A_1 B_1 E_1}{E_1^2 - E^2 - iB_1 E} + \frac{A_2 B_2 E_2}{E_2^2 - E^2 - iB_2 E}. \qquad (S2)$$

Here we have neglected the anisotropy of the dielectric constant. Validity of this approximation is justified by the satisfactory fitting quality for the data measured at different angles of incidence with a single set of the fitting parameters: $A_D = 85.1$, $B_D = 0.497$, $E_1 = 1.38$, $B_1 = 1.64$, $E_2 = 15.6$, , $B_2 = 10.0$ (in unit of eV), with $\varepsilon_\infty$ = -5.131, $A_1 = 10.81$, and $A_2 = 13.7$, as shown in Fig. S1(a). The corresponding complex $\varepsilon(E)$ is deduced and plotted in Fig. S1(b).

With $\varepsilon(E)$ known, we can calculate the tensorial transmission Fresnel coefficients $\overleftrightarrow{L}$ from the previously derived expressions [4, 5]:

$$L_{xx} = \frac{2k_z^{II}}{\varepsilon^{II}k_z^I + k_z^{II}}, \tag{S3a}$$

$$L_{yy} = \frac{2k_z^I}{k_z^I + k_z^{II}}, \tag{S3b}$$

$$L_{zz} = \frac{2\varepsilon^{II}k_z^I}{\varepsilon^{II}k_z^I + k_z^{II}} \cdot \frac{1}{\varepsilon'}, \tag{S3c}$$

where $\varepsilon^\Omega$ is the (bulk) dielectric constant, and $k_z^\Omega$ is the z-component of the wavevector, in medium $\Omega$. ($\Omega = I$ and $II$ for denoting air and the sample, respectively.) $\varepsilon'$ is the dielectric constant of the interfacial region where SHG occurs. We approximate $\varepsilon' \cong \varepsilon^{II}$ [5]. The calculation results for PdTe$_2$ at $\omega$ and $2\omega$ are summarized in Table S1 for the HD-SHG setup configuration and Table S2 for the modified homodyne setup with two input beams.

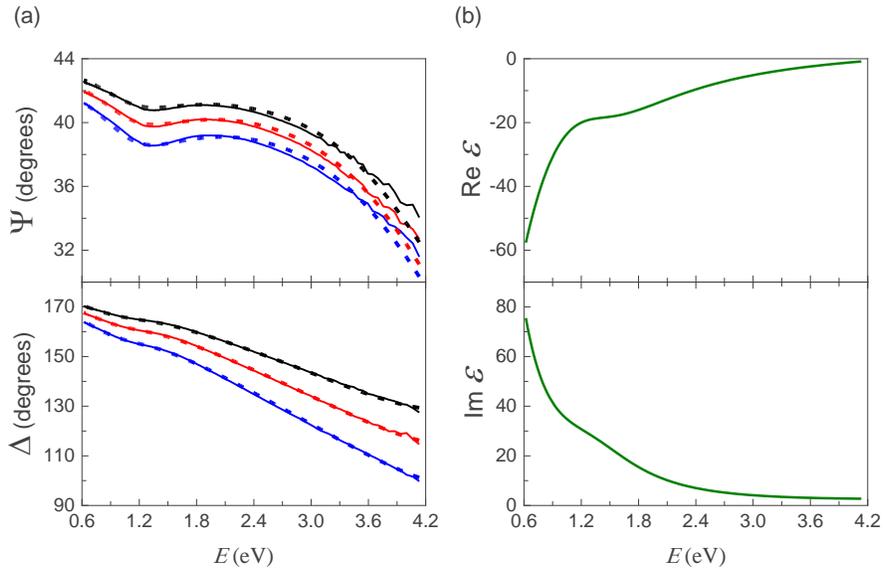

**Figure S1.** (a) Ellipsometry parameters measured from PdTe$_2$ crystal with incidence angles of 50° (black), 55° (red), and 60° (blue) (Solid lines). Dashed lines are the corresponding Drude–Lorentz fits. (b) Complex $\varepsilon(E)$ deduced from the ellipsometric analysis in (a).

| | | |
|---|---|---|
| $L^{\omega}_{xx} = 0.288 - 0.370i$ | $L^{\omega}_{yy} = 0.130 - 0.213i$ | $L^{\omega}_{zz} = -0.028 - 0.055i$ |
| $L^{2\omega}_{xx} = 0.640 - 0.619i$ | $L^{2\omega}_{yy} = 0.266 - 0.422i$ | $L^{2\omega}_{zz} = -0.107 - 0.225i$ |

**Table S1.** Transmission Fresnel coefficients of PdTe$_2$ at $\omega$ and $2\omega$ for the HD-SHG setup

| | | |
|---|---|---|
| $L^{\omega}_{1,xx} = 0.274 - 0.359i$ | $L^{\omega}_{1,yy} = 0.136 - 0.221i$ | $L^{\omega}_{1,zz} = -0.029 - 0.055i$ |
| $L^{\omega}_{2,xx} = 0.303 - 0.383i$ | $L^{\omega}_{2,yy} = 0.123 - 0.204i$ | $L^{\omega}_{2,zz} = -0.028 - 0.055i$ |
| $L^{2\omega}_{xx} = 0.639 - 0.619i$ | $L^{2\omega}_{yy} = 0.267 - 0.422i$ | $L^{2\omega}_{zz} = -0.107 - 0.225i$ |

**Table S2.** Transmission Fresnel coefficients of PdTe$_2$ at $\omega$ and $2\omega$ for the modified homodyne setup with dual input beams. $\overleftrightarrow{L}^{\omega}_i$ with $i = 1,2$ denotes the input at the incident angle $\beta_{Fi}$.

We express $\chi_{eff}$ of the PdTe$_2$ surface in terms of $\overleftrightarrow{\chi}^{(2)}$ for the different setup configurations by Eq. (1) in the main text, where the effects of the $\overleftrightarrow{L}$ and the polarization projection factors are lumped into complex constants $c_i$ ($i = 1 \sim 13$). The expressions of $c_i$ are given as:

$$c_1 = L^{2\omega}_{yy}(L^{\omega}_{yy})^2, \tag{S4a}$$

$$c_2 = -L^{2\omega}_{yy}(L^{\omega}_{xx})^2 \cos^2 \beta_F, \tag{S4b}$$

$$c_3 = L^{2\omega}_{zz}(L^{\omega}_{yy})^2 \sin \beta_{SH}, \tag{S4c}$$

$$c_4 = L^{2\omega}_{xx}(L^{\omega}_{yy})^2 \cos \beta_{SH}, \tag{S4d}$$

$$c_5 = L^{2\omega}_{yy}\left[L^{\omega}_{1,zz}L^{\omega}_{2,yy} \sin \beta_{F1} + L^{\omega}_{1,yy}L^{\omega}_{2,zz} \sin \beta_{F2}\right], \tag{S4e}$$

$$c_6 = -L^{2\omega}_{yy}\left[L^{\omega}_{1,xx}L^{\omega}_{2,yy} \cos \beta_{F1} + L^{\omega}_{1,yy}L^{\omega}_{2,xx} \cos \beta_{F2}\right], \tag{S4f}$$

$$c_7 = L^{2\omega}_{yy}L^{\omega}_{zz}L^{\omega}_{yy} \sin \beta_F \sin 2\psi, \tag{S4g}$$

$$c_8 = -L_{yy}^{2\omega} L_{xx}^{\omega} L_{yy}^{\omega} \cos\beta_F \sin 2\psi, \tag{S4h}$$

$$c_9 = L_{yy}^{2\omega} \left[ (L_{yy}^{\omega})^2 \sin^2\psi - (L_{xx}^{\omega})^2 \cos^2\beta_F \cos^2\psi \right], \tag{S4i}$$

$$c_{10} = -L_{xx}^{2\omega} (L_{xx}^{\omega})^2 \cos\beta_{SH} \cos^2\beta_F \tag{S4j}$$

$$c_{11} = L_{zz}^{2\omega} (L_{xx}^{\omega})^2 \sin\beta_{SH} \cos^2\beta_F, \tag{S4k}$$

$$c_{12} = -L_{xx}^{2\omega} \cos\beta_{SH} L_{xx}(\omega) L_{zz}(\omega) \sin 2\beta_F \tag{S4l}$$

$$c_{13} = L_{zz}^{2\omega} (L_{zz}^{\omega})^2 \sin\beta_{SH} \sin^2\beta_F, \tag{S4m}$$

where $\psi$ in Eq. (S4g) - (S4i) for SXX polarization combination is the angle of the linear polarization of the input beam away from the scattering plane. It was estimated to be ~42° in the experiment.

### S3. Effective surface nonlinear optical susceptibility of quartz reference

We calculate $\chi_{eff}$ of the crystalline $z$-cut quartz reference. To access the dominating terms in its bulk susceptibility $\overleftrightarrow{\chi}_B^{(2)}$ of the quartz, $\chi_{B,xxx}^{(2)} = -\chi_{B,xyy}^{(2)} = -\chi_{B,yyx}^{(2)} = -\chi_{B,yxy}^{(2)} (= \chi_q)$, we set $x$ ($y$) axis of the crystal on the scattering plane in the PSS, PPP, SSP, and SXX (SSS and SPP) experiments. The corresponding $\chi_{eff}$ for quartz with these polarizations is expressed as

$$\chi_{eff}^{SSS} = iL_{yy}^{2\omega} (L_{yy}^{\omega})^2 \chi_q \Delta k_z^{-1}, \tag{S5a}$$

$$\chi_{eff}^{SPP} = -iL_{yy}^{2\omega} (L_{xx}^{\omega})^2 \cos^2\beta_F \, \chi_q \Delta k_z^{-1}, \tag{S5b}$$

$$\chi_{eff}^{PSS} = iL_{xx}^{2\omega} (L_{yy}^{\omega})^2 \cos\beta_{SH} \, \chi_q \Delta k_z^{-1}, \tag{S5c}$$

$$\chi_{eff}^{SSP} = -iL_{yy}^{2\omega} [L_{1,yy}^{\omega} L_{2,xx}^{\omega} \cos\beta_{F2} + L_{2,yy}^{\omega} L_{1,xx}^{\omega} \cos\beta_{F1}] \chi_q \Delta k_z^{-1}, \tag{S5d}$$

$$\chi_{eff}^{SXX} = -iL_{yy}^{2\omega} L_{yy}^{\omega} L_{xx}^{\omega} \cos\beta_F \sin 2\psi \, \chi_q \Delta k_z^{-1}, \tag{S5e}$$

$$\chi_{eff}^{PPP} = -iL_{xx}^{2\omega}(L_{xx}^{\omega})^2 \cos\beta_{SH} \cos^2\beta_F \chi_q \Delta k_z^{-1}, \tag{S5f}$$

where $\Delta k_z$ is the phase mismatch for reflected SHG [4, 5]. We calculate $\chi_{eff}$ by Eq. (S5) with the ordinary refractive index of 1.54 (1.56) and the extraordinary refractive index of 1.55 (1.57) at $\omega$ ($2\omega$), $\chi_q = 2d_{11} = 0.6$ pm/V [6], and $\Delta k_z^{-1} \sim 23$ nm.

## S4. Supplementary figures

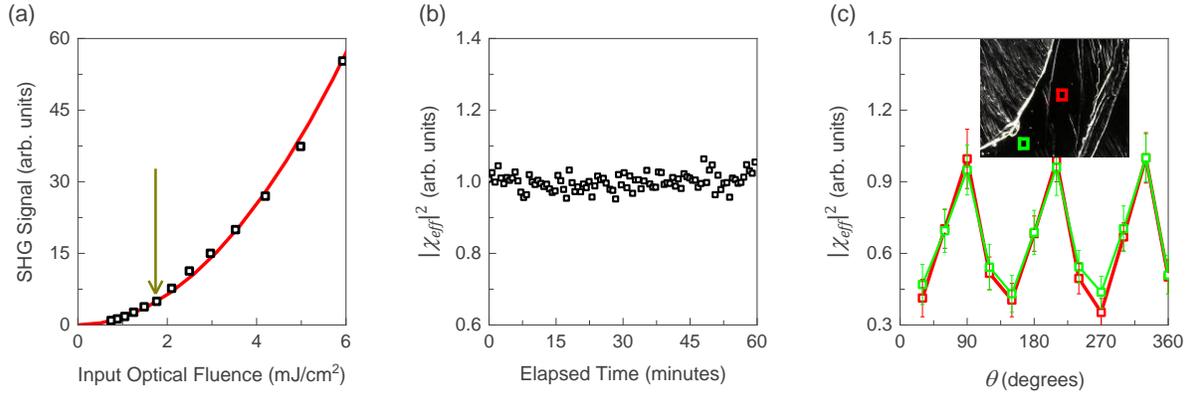

**Figure S2.** (a) SHG signal from PdTe$_2$ versus the input optical fluence (dots). Line is a fit with quadratic dependency with respect to the input fluence. Reasonable fitting quality confirms that the photons detected originate from SHG. An arrow indicates the optical fluence adopted in our HD-SHG study. (b) $|\chi_{eff}|^2$ measured as a function of the laser exposure time for a fully oxidized PdTe$_2$ surface which was stored in dark for >1 week before the measurement. Invariance of the SHG intensity upon the irradiation, together with indiscernible change in the optical microscope image at the location of the laser spot (not shown here), reveals the absence of laser-induced (ir)reversible processes. (c) PPP-polarized RA-SHG intensity measured from two well-separated

locations on the PdTe$_2$ surface, as marked on an optical microscope image of the sample in the inset. Consistency of the data supports good homogeneity of the surface property.

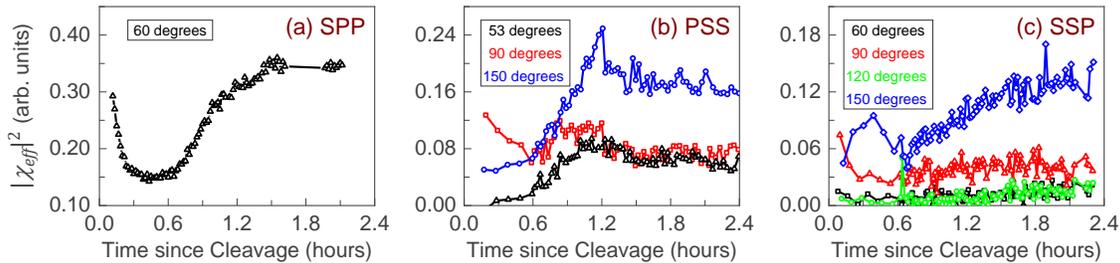

**Figure S3.** Time evolution of the measured $|\chi_{eff}|^2$ from (001) PdTe$_2$ since its surface cleavage for (a) SPP, (b) PSS, and (c) SSP polarizations at selected azimuthal angles.